\documentclass[prl,twocolumn,superscriptaddress,floatfix]{revtex4-2}

\usepackage[latin1]{inputenc}
\usepackage{graphicx}
\usepackage{amsmath}
\usepackage{amssymb}
\usepackage{color}
\usepackage[dvips]{epsfig}
\usepackage{epstopdf}
\usepackage{times,mathptmx}
\usepackage{epstopdf}
\usepackage{natbib}

\DeclareMathOperator{\Ei}{Ei}

\begin{document}
	\title{Comment on ``Conductivity of Concentrated Electrolytes''}

	\author{Olga I. Vinogradova}
	\affiliation{Frumkin Institute of Physical Chemistry and Electrochemistry, Russian Academy of Sciences, 31 Leninsky Prospect, 119071 Moscow, Russia}

\author{Elena F. Silkina}
	\affiliation{Frumkin Institute of Physical Chemistry and Electrochemistry, Russian Academy of Sciences, 31 Leninsky Prospect, 119071 Moscow, Russia}
	
	\maketitle

In their Letter~\cite{avni.y:2022}  Avni \emph{et al.} employed a stochastic density functional theory (SDFT)~\cite{demery.v:2016}
to derive an approximate formula that describes the conductivity $K$ of 1:1 electrolytes of concentration $c_{\infty}$ (with ions of a hydrodynamic radius $R$):
\begin{widetext}
	\centering
	\begin{equation}\label{eq:avni}
		1-\frac{K}{K_0}  \simeq   \varrho e^{-a/\lambda_D}
		+ \dfrac{1}{6} \left( 1 - \frac{1}{\sqrt{2}} + e^{-2a/\lambda_D} - \frac{1}{\sqrt{2}}  e^{-\sqrt{2}a/\lambda_D} \right)\dfrac{\ell_{B}}{\lambda_{D}},
	\end{equation}
\end{widetext}
where $K_0$ is the conductivity at an infinite dilution, $\lambda_D \simeq 0.305 [\rm{nm}]/\sqrt{c_{\infty}[\rm{mol/l}]}$ is the Debye length, $\varrho = R/\lambda_D$,  $a = r_{+} + r_{-} $ is the cut-off length defined as the sum of the cation and anion crystallographic radii,  and $\ell_{B}$ is the Bjerrum length.
 These authors conclude that Eq.~\eqref{eq:avni} significantly improves the Onsager formula~\cite{onsager.l:1927} (reproduced when $a=0$),
and provides a reasonable fit to the conductivity data up to 3 mol/l.
Below we argue that Eq.~\eqref{eq:avni} does not apply at molar concentrations.

\begin{figure}[h]
	\begin{center}
\includegraphics[width=8.6cm]{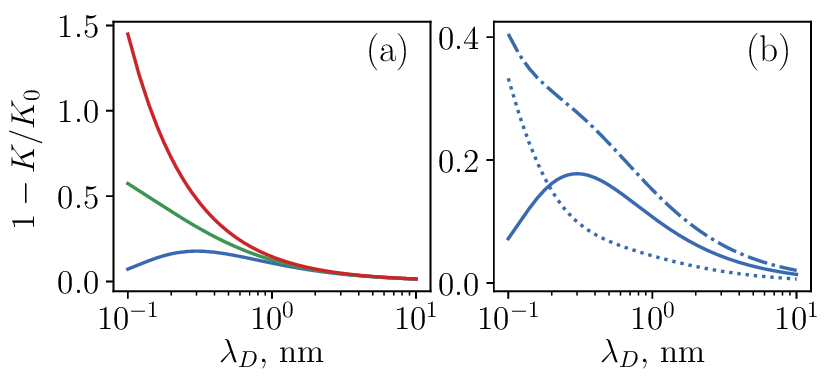}
	\end{center}
	\caption{(a) Electrophoretic term calculated for $R=0.145$ nm  from Eqs.~\eqref{eq:avni} with $a=0$, \eqref{eq:Henry}, and \eqref{eq:avni} with $a=0.3$ nm  (from top to bottom); (b) electrophoretic (solid curve) and relaxation (dotted curve) term and they sum (dash-dotted curve) calculated from \eqref{eq:avni}.}
	\label{fig:K_check}
\end{figure}

The first term
in the r.h.s. of \eqref{eq:avni} is associated with an  ionic electrophoresis under external field. Since the latter takes its origin in the electrostatic diffuse layer (EDL), on increasing $c_{\infty}$ (reducing $\lambda_D$) its speed reduces. Consequently, the electrophoretic term should augment. However, in Eq.~\eqref{eq:avni}
it exhibits a maximum at $\lambda_D=a$ and turns to 0, when $\lambda_D \to 0$ (see Fig.\ref{fig:K_check}(a)). This implies that the conductivity is the same as at an infinite dilution, which is counterintuitive. Since $a \simeq 0.3$ nm~\cite{avni.y:2022}, the electrophoretic term in \eqref{eq:avni} becomes unphysical when $c_{\infty}\geq 1$ mol/l. This term should provide $K/K_0 \to 0$ when $\lambda_D \to 0$ and reduce to the Onsager electrophoretic term if $\varrho \ll 1$, and is given by~\cite{vinogradova.oi:2023b}
\begin{equation}\label{eq:Henry}
 1 - \frac{K}{K_0} \simeq 1 - \frac{3}{2} \frac{\mathcal{F}}{1+\varrho},
\end{equation}
where $\mathcal{F}=1 - e^{\varrho} \left[5 \Ei_7(\varrho) - 2 \Ei_5(\varrho)\right]$~\cite{henry.dc:1931}.

The second term
in \eqref{eq:avni}
is the so-called relaxation, which is due to an emergence of an additional (retarding) field caused by a distortion of the EDL symmetry. The relaxation contribution (i) should vanish at $c_{\infty} \to 0$  and at high salt~\cite{overbeek}: no EDL (of thickness $\lambda_D$) - nothing to distort; (ii) (as a second order effect) is always smaller than electrophoretic. However, this term in \eqref{eq:avni} also demonstrates incorrect trend as seen in Fig.~\ref{fig:K_check}(b). Instead of having  a maximum and then reducing with salt, its magnitude  increases monotonically. When $a=O(\lambda_D)$ this term becomes of the same order as electrophoretic, and in more concentrated solutions even dominates. As a result, when the EDL  nearly disappears its relaxation grows. Clearly, it cannot be so. Also included in Fig.~\ref{fig:K_check}(b) are the electrophoretic contribution and the sum of both terms. It becomes evident that the relative success of \eqref{eq:avni} is simply due to a partial compensation of the incorrect behavior of separate contributions when they are summed up.

In summary, Eq.~\eqref{eq:avni} by \citet{avni.y:2022} derived by including a finite ion size (through $a$) into the SDFT  improves the conductivity description compared to Onsager and extends the range up to \emph{ca.} 1 mol/l, but at larger concentrations leads to physically wrong predictions. Thus these authors  overstated by concluding that their equation applies up to a few mol/l.

O.I.V. and E.F.S. are supported by the Ministry of Science and Higher Education of the Russian Federation.

\bibliographystyle{apsrev4-2}
\bibliography{ion}
\end{document}